\documentclass[onecolumn]{revtex4-1}
\usepackage{graphicx}
\usepackage{caption}
\usepackage{subcaption}
\usepackage{dcolumn}
\usepackage{bm}
\usepackage{xcolor}
\begin{document}


\title{Proton Light Yield in Organic Scintillators using a Double Time-of-Flight Technique}

\author{J.~A.~Brown}
\affiliation{Department of Nuclear Engineering, University of California, Berkeley, California 94720 USA}
\author{B.~L.~Goldblum}
\email{bethany@nuc.berkeley.edu}	
\affiliation{Department of Nuclear Engineering, University of California, Berkeley, California 94720 USA}
\author{T.~A.~Laplace}	
\affiliation{Department of Nuclear Engineering, University of California, Berkeley, California 94720 USA}
\author{K.~P.~Harrig}
\affiliation{Department of Nuclear Engineering, University of California, Berkeley, California 94720 USA}
\author{L. A.~Bernstein}
\affiliation{Department of Nuclear Engineering, University of California, Berkeley, California 94720 USA}
\affiliation{Lawrence Berkeley National Laboratory, Berkeley, California 94720 USA}
\author{D.~L.~Bleuel}
\affiliation{Lawrence Livermore National Laboratory, Livermore, California 94551 USA}
\author{W.~Younes}
\affiliation{Lawrence Livermore National Laboratory, Livermore, California 94551 USA}
\author{D. Reyna}
\affiliation{Sandia National Laboratories, Livermore, California 94550 USA}
\author{E. Brubaker}
\affiliation{Sandia National Laboratories, Livermore, California 94550 USA}
\author{P. Marleau}
\affiliation{Sandia National Laboratories, Livermore, California 94550 USA}
\date{\today}

\begin{abstract}
Recent progress in the development of novel organic scintillators necessitates modern characterization capabilities. As the primary means of energy deposition by neutrons in these materials is n-p elastic scattering, knowledge of the proton light yield is paramount. This work establishes a new model-independent method to continuously measure proton light yield in organic scintillators over a broad energy range. Using a deuteron breakup neutron source at the 88-Inch Cyclotron at Lawrence Berkeley National Laboratory and an array of organic scintillators, the proton light yield of EJ-301 and EJ-309, commercially available organic liquid scintillators from Eljen Technology, were measured via a double time-of-flight technique. The light yield was determined using a kinematically over-constrained system in the proton energy range of $1-20$~MeV. The effect of pulse integration length on the magnitude and shape of the proton light yield relation was also explored. This work enables accurate simulation of the performance of advanced neutron detectors and supports the development of next-generation neutron imaging systems.    
\end{abstract}

                             
                             
\maketitle

\section{Introduction}
Light yield measurements have a long history of providing insight into the response of neutron detection systems. As neutron spectroscopy methods and imaging techniques mature in response to a strong interest in nonproliferation-enabling technologies, the ability to characterize the proton light yield of scintillator materials has become of fundamental importance. For example, neutron imaging techniques using double-scatter kinematic reconstruction depend on the proton light yield relation to determine the energy of the recoil proton following an n-p elastic scattering event and thereby infer the angle of the incident neutron.\cite{brennan, weinfurther} The light yield relation is also used as input to Monte Carlo simulations of the response of neutron detection systems, which in turn is a critical input for neutron spectroscopy techniques using pulse height spectrum unfolding.\cite{miller, lawrence-unfolding} 

Several categories of methods exist for the measurement of proton light yield---direct methods, indirect methods, and edge characterization techniques---as detailed in Sec.~\ref{foundation}. This work introduces a new type of indirect method for measuring the proton light yield of organic scintillators. The approach uses a double time-of-flight (TOF) technique and a broad spectrum neutron source to continuously determine the light yield over a wide range of proton recoil energies without changes to the detector system configuration. Using an array of organic liquid scintillators, the incident neutron energy, known neutron scattering angle, and scattered neutron energy provide a kinematically over-constrained system to determine the proton recoil energy in n-p elastic scattering events. This double TOF approach, described in Sec.~\ref{method}, obviates the uncertainty associated with characterizing the maximum energy-transfer edge in proton recoil pulse height spectra and provides strong background rejection criteria for neutron-carbon interactions and double scatters (i.e., two neutron interactions within the scintillator). Section~\ref{exp} describes the experimental setup for a demonstration of the method via proton light yield measurements of EJ-301 and EJ-309, organic liquid scintillators manufactured by Eljen Technology with pulse shape discrimination properties. The EJ-301 scintillator was examined as the body of early literature on light yield measurements was highly focused on the commercial equivalent: NE-213 from Nuclear Enterprise.\cite{verbinski} With a similar formulation to EJ-301, a higher flash point, and a lower chemical toxicity, EJ-309 is quickly becoming an industry standard and has been targeted extensively in recent light yield measurements.\cite{pino, enqvist, takada, lawrence} The data analysis and uncertainty quantification procedures for the EJ-301 and EJ-309 proton light yield measurements are detailed in Sec.~\ref{analysis}. The results are given in Sec.~\ref{results}, along with an evaluation of the effect of pulse integration length. Concluding remarks are provided in Sec.~\ref{conc}.


\section{Foundational Work}
\label{foundation}

The measurement and characterization of scintillator light yield has been accomplished through a variety of methods that fall into three main categories. Direct methods involve charged particle beams made incident on a scintillator to obtain a direct determination of the energy deposited in the medium. Indirect methods use n-p elastic scattering kinematics and coincident neutron measurements between two or more detectors to determine the proton recoil energy deposited in a scintillator. Edge characterization techniques rely on the understanding that a neutron can transfer up to all of its energy in a single elastic collision with a proton. As such, a measurement of the response of a scintillator to a series of monoenergetic or quasi-monoenergetic incident neutron fluxes is conducted, where the maximum edge of the pulse height (or pulse integral) distribution corresponds to the energy of the incident neutron. Each of these approaches is explored in more detail below.

\subsection{Direct Methods}
In direct approaches to the measurement of light yield for organic scintillators, energetic charged particles are impinged directly on the scintillating medium of interest. One of the earliest examples of this technique comes from Franzen et al.\ in 1950.~\cite{Franzen} This experiment used a 16.4~MeV proton beam and a series of degrader foils to cover the energy range of interest, where the peak pulse height was used to characterize the light output. More recently, Takada et al.\ used energetic protons up to $70$~MeV to measure the proton light yield of an EJ-309 organic liquid scintillator in a phoswich-type configuration.\cite{takada} A thorough review of the early direct measurements is provided by Birks.~\cite{birks}  

The direct approach to light yield measurement poses several challenges. The energy deposition profile of the impinging particles is shallow with respect to the incident direction of the charged particle beam, necessitating the use of thin samples. The relative uncertainty in the charged particle energy deposition increases with decreasing energy due to energy straggling (and energy loss in the detector housing in the case of liquid scintillators). Further, the scintillator must be coupled to the vacuum system of the radiation-producing machine, which for liquid scintillators requires the development of specialized detector housings. 

\subsection{Indirect Methods}

Indirect approaches to measuring proton light yield involve determination of the recoil proton energy using neutron coincidence measurements and n-p elastic scattering kinematics. Pioneering work by Smith et al.\ used a set of two detectors and a series of monoenergetic neutron sources to measure coincident events between the two detectors.\cite{SMITH1968157} To cover a broad energy range, Smith et al.\ moved the secondary detector to various angles throughout the experiment and the proton recoil energy was obtained using the known incident neutron energy and scattering angle. 

Several extensions of this method can be found in later work. Galloway and Savalooni used an indirect approach to the light yield measurement of NE-213 with an AmBe source.\cite{GALLOWAY1982549} In this case, neutrons were made incident on the scintillator and the energy of the scattered neutron was obtained via TOF to a secondary detector. The recoil proton energy was then calculated using n-p elastic scattering kinematics, the scattered neutron energy, and the known scattering angle. The KamLAND collaboration used multiple secondary detectors (in fixed position) and a mono-energetic neutron source to measure the light yield using the known incident neutron energy and scattering angle.\cite{Yoshida2010574} Though exceptions exist, work on indirect methods has relied primarily on monoenergetic incident neutron fields. 

When using a monoenergetic neutron source, any given detector pair leads to a single proton recoil energy in the primary detector. To obtain the light yield relation across the energy range of interest, such an approach requires repositioning of the secondary detectors, a large number of secondary detectors, and/or re-tuning of the incident neutron source. Working with broad spectrum sources (e.g., AmBe, PuBe, etc.) and outgoing TOF allow for a continuous measurement of the light yield relation. However, such sources pose experimental challenges due to efficiency limitations of the coincident detection array. Efforts to mitigate low detection efficiency through the use of shorter flight paths results in increased uncertainty in the TOF measurement.   

\subsection{Edge Characterization Methods}

For an ideal detector with infinite resolution and a perfectly linear light yield, the proton light response to a monoenergetic neutron source would be a rectangular function with a maximum edge at the incident neutron energy. This is due to the isotropic angular n-p scattering cross section in the center-of-mass system, which leads to an equal probability of transferring any energy to the recoil proton up to the incident neutron energy. Edge characterization methods exploit this idealization of the response function by measuring monoenergetic (or quasi-monoenergetic) pulse height or pulse integral spectra at multiple energies and relating the proton recoil edge to the incident neutron energy. 

A variety of methods have been employed to characterize the position of an edge, which in practice is smeared by the energy resolution of the detection system.\cite{scherzinger, brown-stilbene} Early work by Verbinski et al.\ involved the collection of a series of monoenergetic response functions, where the half-height of the edge of the distribution was related to the incident neutron energy.\cite{verbinski} Initial measured values were taken as a trial light yield relation, which was used as input for a Monte Carlo simulation of detector response functions. The trial light yield relation was then modified until agreement was obtained between the simulated and measured response functions. More recent work by Kornilov et al.\ used the first derivative of the response function to characterize the proton recoil edge.~\cite{Kornilov2009226} Instead of using a fractional height of the edge, the data were smoothed and the first derivative was estimated numerically leading to an inverted normal distribution centered at the idealized edge. The normal distribution was then fit, with the centroid taken as the position of the edge. Recently, Scherzinger et al.\ compared a raw set of light yield data using a half-height position (without a trial light yield simulation), a first derivative edge, and a technique using the turning point of a Gaussian function fit to the resolution-smeared proton recoil edge.~\cite{scherzinger} Scherzinger compared resultant Monte Carlo simulations to measured response functions for 3, 4, and 5 MeV neutrons and noted that the half-height method provided the best result. In general, the Scherzinger simulation results disagreed with the data not only in absolute magnitude, but included a bias as a function of energy with discrepancies between the three methods increasing with decreasing neutron energy.

In practice, the response functions of neutron detection systems deviate from idealized step functions. This difference arises due to effects from multiple scatters, protons escaping through the detector cell walls, neutron-carbon interactions, light yield non-linearity, and the energy dependence of the resolution function.\cite{knoll2010radiation} The possible bias of these effects on edge characterization approaches to light yield measurement is not extensively explored in the literature. If the maximum energy-transfer edge is used as a starting point for a simulation which accounts for these distortions and iterates on the light yield relation, it may be possible to obtain the relation using edge characterization without bias.     


\section{Methodology}
\label{method}

This work outlines a double TOF technique for measuring light yield using an experimental setup akin to those employed in indirect approaches. Instead of using a mono-energetic incident source, a pulsed high-flux broad spectrum neutron beam is used as the source. Target detectors, comprised of the scintillator to be characterized, are configured in a collimated beam with one or more scattering detectors placed outside the primary neutron beam at forward angles relative to the axis defined by the incoming beam. This provides a kinematically over-constrained system that allows for TOF measurements of both the incident and scattered neutron.   

Using the neutron TOF method, the relativistic energy-time relationship is employed to determine both the energy of the incident and scattered neutron, $E_n$ and $E^{\prime}_n$, respectively. Here, 
\begin{equation}
\label{E_TOF}
E_n = (\gamma - 1)m_n c^2,
\end{equation}
where the Lorentz factor, $\gamma$, is given by:
\begin{equation}
\label{gamma}
\gamma = \frac{1}{\sqrt{1-\frac{(d/t)^2}{c^2}}}.
\end{equation}
Here, $m_n$ is the rest mass of the neutron, $c$ is the speed of light, $d$ is the neutron flight path, and $t$ is the TOF of the neutron. For the scattered neutron, the TOF is measured between the target and the scattering detectors. The scattering angle of the neutron in the forward direction, $\theta$, is known from the geometry of the detector array configuration. 

Using n-p elastic scattering kinematics, the proton recoil energy, $E_p$, can then be determined via the incoming and outgoing neutron energies: 
\begin{equation}
E_p = E_n - E^{\prime}_n,
\end{equation}
the incoming neutron energy, $E_n$, and the scattering angle, $\theta$:
\begin{equation}
\label{thisone}
E_p = E_n \sin^2\!{\theta}, 
\end{equation}
and the outgoing neutron energy, $E^{\prime}_n$, and the scattering angle, $\theta$:
\begin{equation}
E_p = E^{\prime}_n \tan^2\!{\theta}. 
\end{equation}
For a given detector configuration, the three different approaches can be evaluated allowing for selection of the method that provides the lowest proton recoil energy uncertainty.\cite{brown-thesis} Consistency among the three approaches can be used as rejection criteria for events that are not n-p elastic scattering. The outgoing TOF can further be used for disambiguation of the incident beam pulse in cases where the duty cycle of the source results in frame overlap.\cite{harrig}

This approach extends work on indirect methods to allow for a continuous light yield measurement over a broad range of energies. Unlike edge characterization techniques that in best practice use feedback between the experimental data and a Monte Carlo simulation, this approach provides a model-independent pathway to the quantity of interest. The measurable proton recoil energy range is limited by the energy distribution of the incoming neutron beam, the angular configuration of the experimental setup, and the dynamic range of the data acquisition system. Multiple scattering angles can be employed to sample different regions of the proton recoil spectrum allowing focus on specific proton recoil energies. If there is a kinematic overlap, where two or more secondary detectors observe the same proton recoil energy range, the efficiency for detection in the overlapping region will increase. This also provides additional confidence in the measurement in that range, both due to increased statistics and the ability to obtain the same result from independent secondary detectors as a systematic check. The high-flux source enables rapid assessment of the proton light yield relation and the use of long flight paths, thereby decreasing the uncertainty in the TOF assessment and uncertainty in the scattering angle while maintaining reasonable event rates. 


\section{Experimental Configuration}
\label{exp}
\begin{figure}
\includegraphics[width=0.47\textwidth]{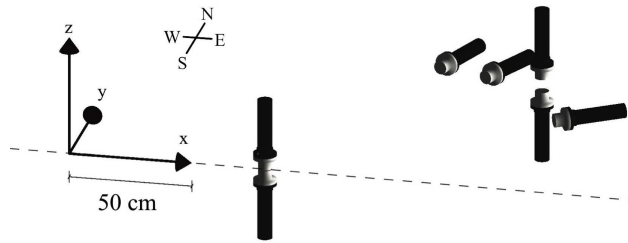}
\caption{\label{scatArrayFig} Geant4 schematic of the detector array configuration. The origin of the coordinate system was positioned at beam line center on the west wall of the experimental area. The neutron beam travels along the $x$-axis (illustrated by the dashed line) through two vertical target cells. The five scattering cells were positioned outside of the neutron beam to observe coincident events.}
\end{figure}

A 33-MeV $^2$H$^+$ beam from the 88--Inch Cyclotron at Lawrence Berkeley National Laboratory was employed to generate breakup neutrons via the $^{nat}$Ta($d,np$) reaction on a $3.8$-cm-thick tantalum target located in the cyclotron vault. The experimental setup was comprised of an array of liquid organic scintillators. A schematic of the detector array was produced using the Geant4 simulation platform and is shown in Fig.~\ref{scatArrayFig}.~\cite{Geant4} Two target detectors were positioned in-beam composed of 5.08 cm dia.\ $\times$ 5.08 cm len.\ right circular cylindrical cells of EJ-301 and EJ-309 oriented up and down, respectively. Five 5.08 cm dia.\ $\times$ 5.08 cm len.\ right circular cylindrical cells (five EJ-309 and one EJ-301) were positioned at forward angles out of beam to observe scattered neutrons. The detector housings were constructed of Al and coupled to Hamamatsu 1949-50 photomultiplier tubes (PMTs) through a thin acrylic window and a thin layer of BC-630 silicone optical grease. 
  
Table~\ref{locations} provides a summary of the measured detector positions and the associated uncertainty. A Cartesian coordinate system was used to reference detector locations with the origin positioned at beamline center on the west wall of the experimental area, which defines the $x=0$ plane. The $x$-coordinate was taken as the distance from the west wall, which was located $647.3 \pm0.5$~cm from the breakup target, the $y$-coordinate was taken as the distance north from beamline center, and the $z$-coordinate was taken as the distance above beamline center. To determine detector locations within this coordinate system, a bi-plane laser was aligned with beamline center. For the $x$ and $y$ measurements, the locations of all detectors were projected onto the floor using a plumb bob. A single plane laser in conjunction with a square was used to create a line from the detector location to beamline center. The length along this line was taken as the $y$-coordinate of a given detector, and the distance along the beamline to the point of intersection was taken as the $x$-coordinate. All detector measurements were made to the center of the front face of the aluminum housing, whereas calculations of flight path and angle were made using the center point of the scintillator cells.

\begin{table} 
\renewcommand{\arraystretch}{1.2}
\begin{tabular}{ccccc}
\hline
Orientation	&	Material	&		x (cm)			&	y (cm)			&	z (cm)		\\ \hline
up	&	EJ-301	&	79.8	$\pm$	0.25	& $	0.0	\pm	0.1	$&$	0.2	\pm	0.1	$\\ 
down 		&	EJ-309	&	79.8	$\pm$	0.25	& $	0.0	\pm	0.1	$&$	-0.1	\pm	0.1	$\\ 
horizontal	&	EJ-309	&	124.3	$\pm$	1	&$	125.1	\pm	1	$&$	0	\pm	0.1	$\\ 
horizontal	&	EJ-309	&	144.2 $\pm$	1	&$	118.3	\pm	1	$&$	0	\pm	0.1	$\\ 
down	&	EJ-309	&	169.4	$\pm$	1	&$	108.55	\pm	1	$&$	5.6	\pm	0.25	$\\ 
up		&	EJ-301	&	169.4	$\pm$	1	&$	108.55	\pm	1	$&$	0	\pm	0.25	$\\ 
horizontal	&	EJ-309	&	181.1 $\pm$	1	&$	76.9	\pm	1	$&$	0	\pm	0.1	$\\ 
\hline
\end{tabular}
\caption{Summary of detector locations and estimated uncertainties in the measurements. \label{locations}}
\end{table}
\par
Data were acquired over a period of approximately 14~h with a beam current of approximately 150~nA. The data were recorded using a CAEN V1730 500 MS/s digitizer set to trigger with one or more events in the target cells and one or more events in the observation cells. The RF control signal from the cyclotron was set to trigger when an event from both a target cell and an observation cell occurred. Full waveforms of the events were written to disk in list mode with global time stamps for pulse processing and event reconstruction in post-processing.
\par
The target PMTs were powered using a CAEN NDT1470 high voltage power supply. The bias was selected to roughly gain match the detectors by setting the full height of signals corresponding to the Compton edge of the $4.4$~MeV $\gamma$~ray from an AmBe source at 25\% of the 2~V full scale range of the digitizer. The PMTs for the scattering detectors were powered using both CAEN~NDT1470 and CAEN~N472 high voltage power supplies. Their gains were set to accommodate pulse heights corresponding roughly to the maximum proton recoil energy anticipated for a given scattering angle.   

\section{Data Reduction and Error Analysis}
\label{analysis}
Data reduction was accomplished in an object-oriented C\texttt{++} framework, which used elements of the ROOT data analysis framework.~\cite{brun} The arrival time of the signals was established using the CAEN digital constant fraction discrimination algorithm, with a 75$\%$ fraction and 4~ns delay for the scintillator signals and a 50$\%$ fraction and 56~ns delay for the cyclotron RF signal.~\cite{DPP-PSD} Waveforms from the scintillator cells were reduced to pulse integrals with 30~ns and 300~ns integration lengths. The 30~ns and 300~ns integration lengths were chosen as these were shown via an evaluation of average waveform data to encompass 75\% and 95\% of the scintillation light, respectively. Although the terms \textit{pulse height} and \textit{pulse integral} are often used interchangeably in the literature,\cite{lawrence} pulse integral is used exclusively herein to differentiate from the pulse height of the digital waveform, which would be proportional to the maximum photon flux, not the number of photons in the scintillation event. The non-linearity of the PMT response functions was measured using a finite difference method adapted from Friend et al.\ and waveforms were corrected on a per sample and per target scintillator basis.~\cite{friendLinearity}  

\begin{figure}
\centering
\includegraphics[width=0.7\textwidth]{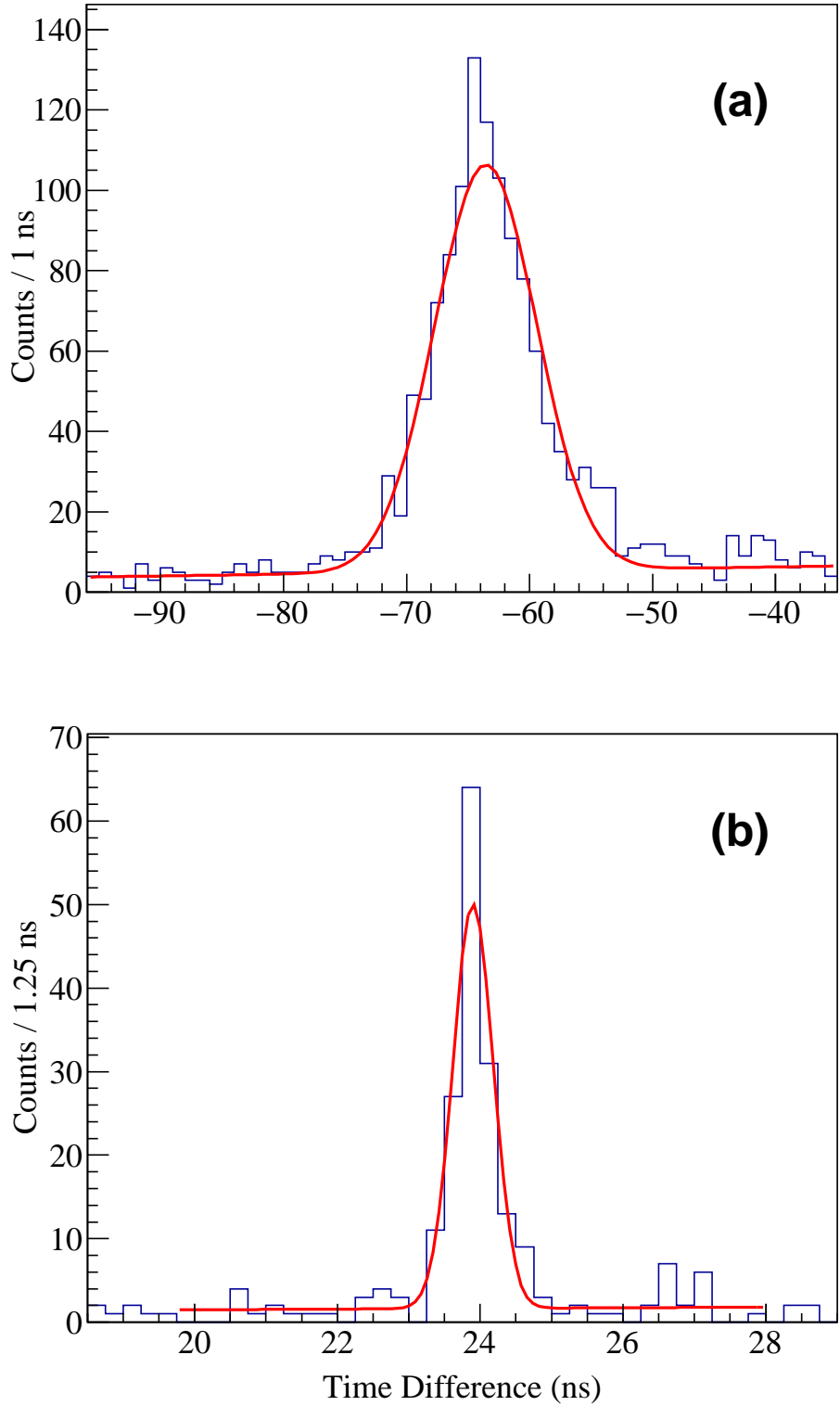}
\caption{The blue curve shows a histogram of time differences between (a) the cyclotron RF signal and $\gamma$-ray events in a target detector and (b) $\gamma$-ray events in the target and observation detectors. The width of the pulse is primarily reflective of (a) the spatial spreading of the beam pulse itself, with a $\sigma=3.94$~ns, and (b) the system time resolution, with a $\sigma=0.25$~ns. The red curve is a fit of the measured data with a normal distribution plus linear background term.}
\label{timeCalib}
\end{figure}

Timing calibrations were performed for both the incident and outgoing TOF measurements. The incident TOF was determined as the time difference between an event in the target cell and the cyclotron RF signal. The measured time, $t_{m1}$, can be represented as: 
\begin{equation}
t_{m1} = t_1 - t_{\mathrm{RF}},
\end{equation}
where $t_1$ is the clock time of the event in the target cell and $t_\mathrm{RF}$ is the time corresponding to the peak of the cyclotron RF signal. To calibrate the signal chain such that the measured time corresponded to the transit time of the particle, a histogram of the arrival time of incident $\gamma$ rays, isolated using pulse shape discrimination in the target cells, was constructed. This resulted in a peak corresponding to the photon flash produced when deuterons impinge on the breakup target. A binned maximum likelihood estimation between this feature and the superposition of a normal distribution with a linear background was performed to obtain the centroid of the distribution, $\bar{x}$, and the associated statistical error. The time calibration constant, $t_c$, is then given by: 
\begin{equation} \label{timeConstant}
t_c = \bar{x} - \frac{d}{c},
\end{equation}  
where $d$ is the flight path and $c$ is the speed of light. A characteristic result of the binned maximum likelihood estimation is shown in Fig.~\ref{timeCalib}a. 

There is an ambiguity in the incoming flight time due to frame overlap, where the cyclotron pulse period, $T_{cyc}$, is shorter than the TOF for incident neutron energies $<23.2$~MeV. In this case, $T_{cyc}=111.094$~ns while flight times range from $\approx80 - 400$~ns. With an integer ambiguity, $n$, the incoming TOF, $t_{inc}$, can be written as: 
\begin{equation}\label{incTOF}
t_{inc} = t_{m1}-t_c+n\times T_{cyc}.
\end{equation}
The uncertainty in the incoming TOF determination is dominated by the uncertainty in $t_c$ with the spreading of the photon flash being $\sigma=3.94$~ns. 

The temporal calibration of the exit TOF was performed using $\gamma-\gamma$ coincidences between the target and observation detectors in a procedure similar to that described above. The measured outgoing TOF, $t_{m2}$, was taken as: 
\begin{equation}
t_{m2} = t_2-t_1,
\end{equation}
where $t_2$ is the clock time of the event in the scattering cell. A calibration time constant was determined by creating a histogram of the $\gamma-\gamma$ coincidences and fitting using the procedure described above. The outgoing TOF is then given by: 
\begin{equation}
t_{out} = t_{m2} - t_c.
\end{equation}
A characteristic fit of the $\gamma-\gamma$ coincidences between the target and observation detectors is shown in Fig.~\ref{timeCalib}b.

\begin{figure*}
\centering
\includegraphics[width=\textwidth]{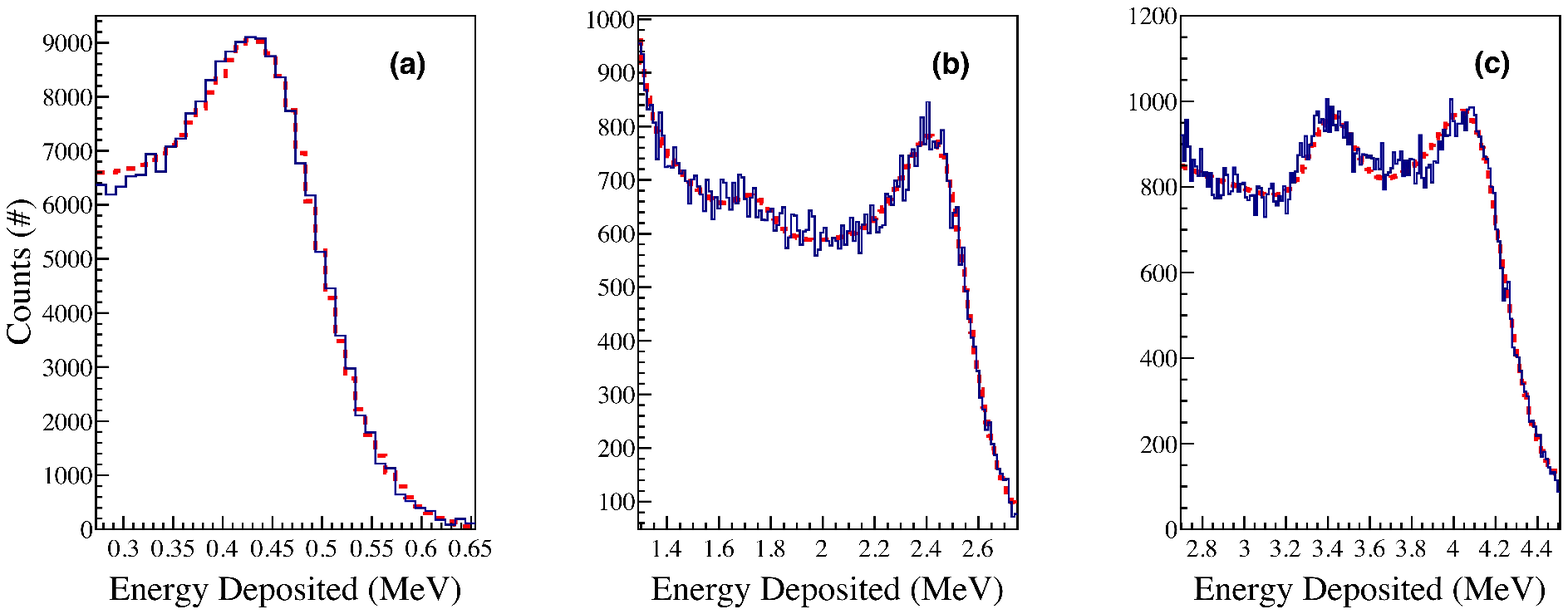}
\caption{Minimization between GEANT4 modeled (red) and experimental source spectra (blue) for (a) the $0.667$~MeV $\gamma$~ray from $^{137}$Cs, (b) the $1.4$~MeV and $2.7$~MeV $\gamma$~rays from aluminum activation, and (c) the $4.44$~MeV $\gamma$~ray from an AmBe source in the EJ-309 target detector. The features at $1.68$~MeV from aluminum activation in (b) and at $3.42$~MeV in the AmBe source spectrum in (c) correspond to the double escape peaks of the $2.7$~MeV and $4.44$~MeV $\gamma$~rays, respectively.}
\label{phCalibFig}
\end{figure*}

To resolve the integer ambiguity, $n$, in the incoming TOF described in Eq.~\ref{incTOF}, a comparison was made between the measured incoming TOF and expected incoming TOF calculated from the scattering angle and the exit energy of the neutron. The expected incoming TOF, $t_{e1}$, is given as 
\begin{equation} \label{expTOF}
t_{e1} = \frac{d_{n^{\prime}}}{c\sqrt{1-\frac{m_n}{m_n+E_{n^{\prime}}}}},
\end{equation}
where $d_{n^{\prime}}$ is the outgoing flight path and $E_{n^{\prime}}$ is the energy of the scattered neutron. To determine the value of $n$, the difference between the measured and expected TOF was divided by the cyclotron period:
\begin{equation}
n = \left|\frac{t_{e1}-t_{m1}}{T_{cyc}}\right|,
\end{equation}
and the nearest whole integer was used. 

For this work, the proton recoil energy was calculated using the incoming TOF and scattering angle as described in Eq.~\ref{thisone}. Using a Monte Carlo simulation that included the relevant sources of uncertainty, this approach to the proton recoil energy determination was shown to provide the lowest uncertainty when considering the full energy range of the assessment.\cite{brown-thesis} The dominant contributor to the proton recoil uncertainty in this configuration was the uncertainty in the incoming TOF, which was dominated by the beam pulse spreading of the cyclotron as shown in Fig.~\ref{timeCalib}a. Note that different experimental configurations and beam pulse spreading may result in alternate optimal approaches to proton recoil energy determination.

The pulse integral was calibrated to an electron-equivalent energy scale, where the scintillation light unit represents the equivalent light for an electron deposition of that energy. Pulse integral spectra were taken with an AmBe source, self-activation from $^{27}$Al$(n,\alpha)$ in the detector housing and beam box, and a $^{137}$Cs source. The AmBe source provided a 4.44~MeV $\gamma$ ray from $^{12}$C de-excitation following the $^9$Be($\alpha$,n) reaction. The intense background from Al activation was the result of a previous experiment that employed high beam currents, which led to two $\gamma$ rays at 2.7~MeV and 1.4~MeV. The $^{137}$Cs source provided a 0.662~MeV $\gamma$ ray. The $^{137}$Cs source spectrum also contained a strong background from the Al activation, so the Al spectrum was subtracted such that the integral above the anticipated Compton edge was zero. Simulations of the energy deposition spectra from these four $\gamma$~rays were performed using the Geant4 framework.\cite{Geant4} To account for the bi-energetic signature from Al, the simulations of the two $\gamma$~rays were performed independently using the same number of particles. These were then combined in proportion to the branching ratio of the decay. 

\begin{table*}
\centering
\renewcommand{\arraystretch}{1.2}
\begin{tabular}{ccccc}
\hline
 & \multicolumn{2}{c}{300 ns Integration} & \multicolumn{2}{c}{30 ns Integration}   \\
Parameter & EJ-301 & EJ-309 & EJ-301 & EJ-309 \\ 
\hline
a                &$ 0.000186               $&$ 0.000217              $ &$ 0.00023               $&$ 0.00026              $\\
b                &$ 0.033                  $&$ 0.0336                $&$ 0.023                  $&$ 0.024                $ \\ 
$\sigma_{a}^2$   &$ 3.8 \times 10^{-17}    $&$ 3.8 \times 10^{-18}   $&$ 5.6 \times 10^{-16}    $&$ 3.12 \times 10^{-16}   $ \\ 
$\sigma_{b}^2$   &$ 2.3 \times 10^{-9}     $&$ 5.02 \times 10^{-11}  $&$ 1.0 \times 10^{-8}     $&$ 6.3 \times 10^{-8}  $ \\ 
$\sigma_{ab}$    &$ -2.24 \times 10 ^{-15} $&$ -1.72 \times 10 ^{-15}$&$ -1.84 \times 10 ^{-13} $&$ -4.4 \times 10 ^{-12}$ \\ 
\hline
\end{tabular}
\caption{Summary of pulse integral calibration results and statistical error for 30~ns and 300~ns integration lengths for the EJ-301 and EJ-309 target detectors.}
\label{calibTable}
\end{table*}

To obtain a linear light output calibration:
 \begin{equation}
L = ax +b,
\end{equation}
where $L$ is the scintillation light in MeV electron-equivalent (MeVee), $x$ is the pulse integral in channels, and $a$ and $b$ are fitting parameters, the measured pulse integral spectra were transformed by $a$ and $b$. The simulation results were scaled and folded with a resolution function of the form:
\begin{equation}
\frac{\delta L}{L} = \sqrt{E_c +\frac{ E_1}{L} + \frac{E_2}{L^2}},
\end{equation}
where $E_c$, $E_1$, and $E_2$ are fitting parameters.~\cite{DIETZE}  A power law:
\begin{equation}
L = c_1 x^{c_2},
\end{equation}
was added to address background contributions, where $c_1$ and $c_2$ are fitting parameters. The model parameters were estimated by minimizing $\chi^2$ between all measured and modeled pulse integral spectra simultaneously. The range for the minimization was truncated as shown in Fig.~\ref{phCalibFig}, where contributions from the 1.4~MeV $\gamma$ ray from aluminum activation were largely neglected due to potential summation from cascade coincidences not accounted for in the simulation. The parameter minimizations were performed using the SIMPLEX and MIGRAD algorithms from the ROOT Minuit2 package, and error matrices were obtained using the HESSE algorithm.\cite{brun} The resultant fit for the pulse integral source data from the EJ-309 target detector calibration is shown in Fig.~\ref{phCalibFig}. The calibration results and error matrix are summarized in Table~\ref{calibTable}. The reported errors include only statistical contributions to the uncertainty and are not representative of the systematic errors in the calibration procedure. The fit range and source locations were varied with a series of calibration sources to estimate the systematic uncertainty on the fit parameters and differences on the order of 1\% were observed. 

The system proton energy resolution is determined by the spatial spread of the deuteron beam leading to an intrinsic temporal uncertainty in the incoming TOF, the angular dispersion due to the spatial extent of the detector, and the flight path variations due to the stochastic interaction location in the detector. Since the angular dispersion varies as a function of the target-observation cell pair, the data were reduced into a two-dimensional histogram of light yield versus proton recoil energy, where the proton energy discretization corresponds to the mean system resolution determined via a Monte Carlo simulation. Additionally, a series of constraints were applied to the data to ensure events included in the histogram arose from single n-p elastic scattering events in the target cells. These include pulse shape discrimination to isolate neutron events, minimum and maximum outgoing neutron energies reflective of the beam energies and observed features, and a requirement that the absolute time difference of the expected and measured incoming TOF was less than 15~ns.\cite{brown-thesis}

\begin{figure}
\includegraphics[width=0.8\textwidth]{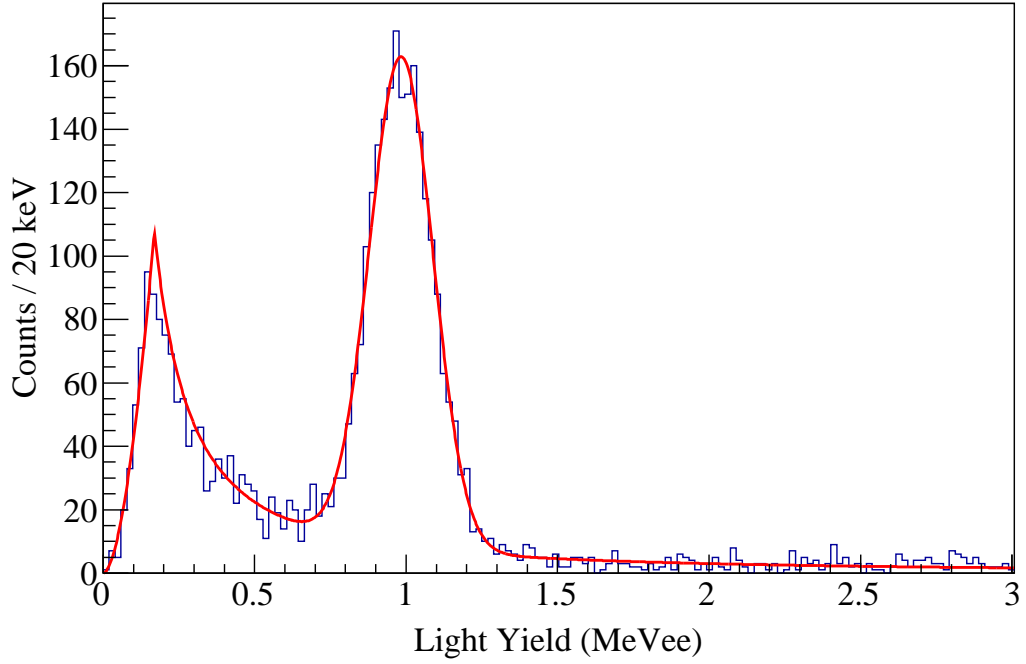}
\caption{Light yield spectrum for a proton energy bin corresponding to proton recoil energies between $2.64-2.93$~MeV fit with a piecewise power law and a normal distribution. The centroid of the normal distribution corresponds to the mean light production for n-p elastic scattering events within this bin.}
\label{sliceFitPlot}
\end{figure}

\begin{figure}
\centering
\includegraphics[width=0.8\textwidth]{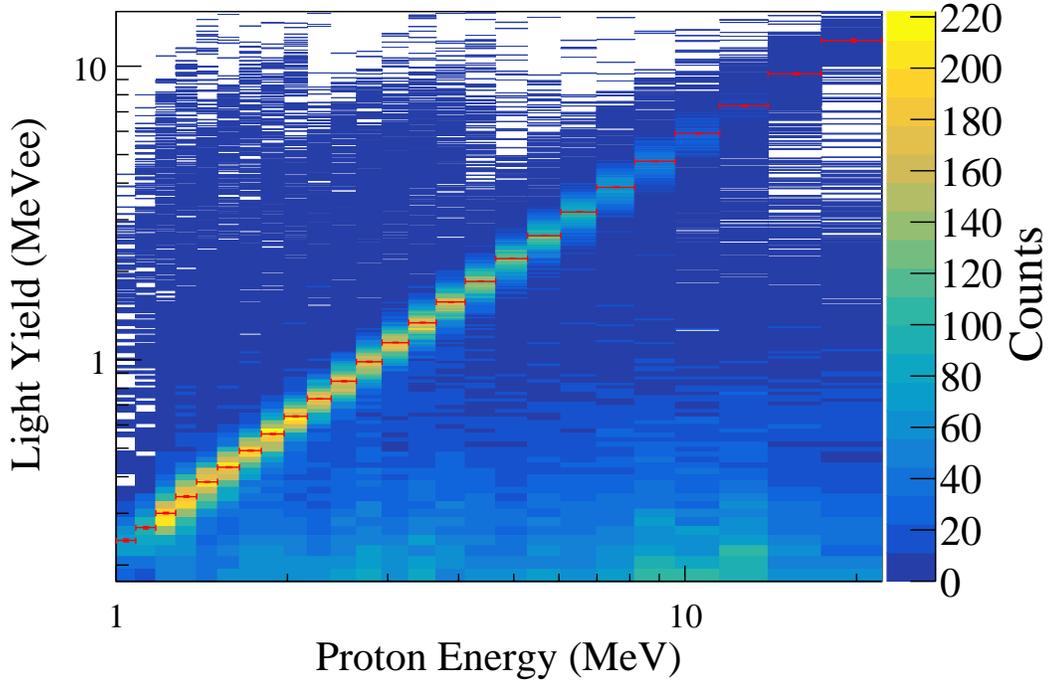}
\caption{Histogram representation of the EJ-309 light yield relation obtained using a 300~ns integration length along with the result from fitting individual proton energy bins with a piecewise defined power law and a normal distribution. The error bars on the ordinate and abscissa values are representative of the uncertainty of the centroid of the n-p elastic scattering feature and the proton recoil energies contributing to the estimate, respectively.}
\label{fittedLyHistFig}
\end{figure}

To produce a series of data points, projections of individual proton energy bins were made on the light yield axis and the resulting histogram was fit using a binned maximum likelihood estimation considering the peak to be normally distributed and the background to be represented by a piecewise power law. A characteristic fit is shown in Fig.~\ref{sliceFitPlot}.
The result is a series of data points where the horizontal error bars represent the proton energies contributing to the estimate, and the vertical error bars are the statistical error on the determination of the centroid of the distribution corresponding to n-p elastic scattering events. This is shown in Fig.~\ref{fittedLyHistFig} along with a histogram representation of the measured EJ-309 proton light yield.

The sources of systematic uncertainty in the measurement include uncertainty in the determination of the incoming and coincident TOF temporal calibration constants, uncertainty in the detector locations, uncertainty in the measured distance from the breakup target to the west wall of the experimental cave, and uncertainty in the light output calibration. To address the uncertainty from these potential sources of bias, a Monte Carlo method was implemented where the reduction from histogrammed quantities to data points was repeated while the input to calculations of the proton energy and light yield were varied. The only correlated variables identified above were the parameters from the light output calibration fitting procedure. The random parameters for these variables were generated through application of Cholesky decomposition (i.e., the product of the lower triangular decomposition of the error matrix and standard normal variates represent the change in the parameters from the best estimate).\cite{recipes} 

\begin{figure}
\centering
\includegraphics[width=0.8\textwidth]{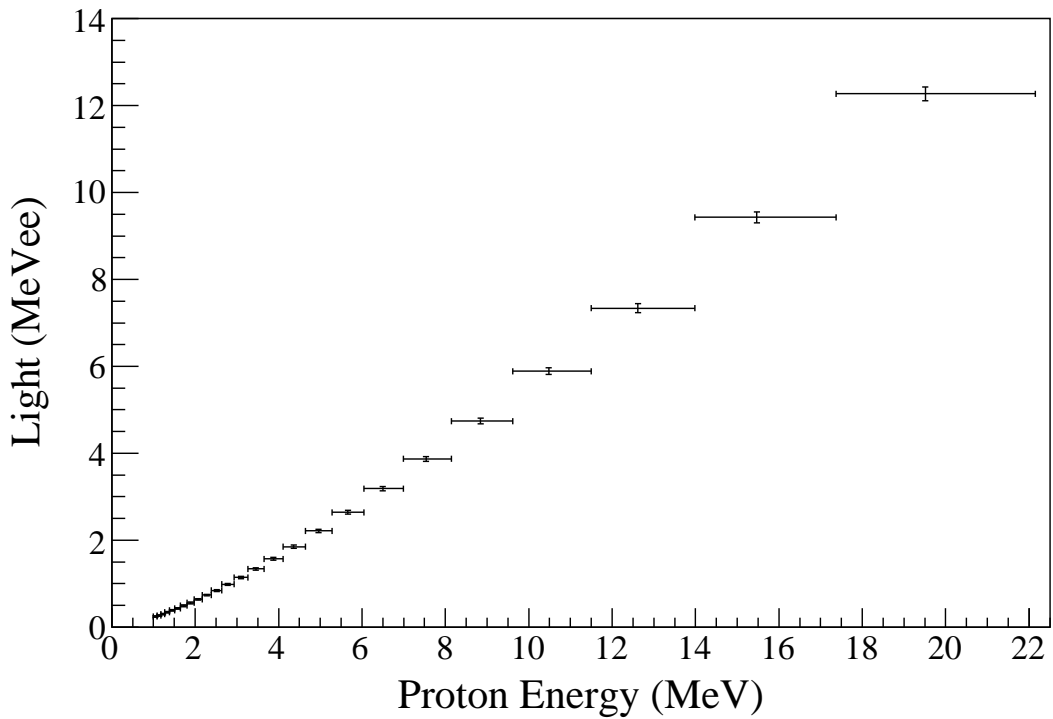}
\caption{Proton light yield of EJ-309 using a 300~ns integration length. The vertical error bars represent the systematic uncertainty in the proton light yield measurements.}
\label{systematicResult}
\end{figure}

The resultant event-wise data points from each trial were stored on disk, histogrammed, and pulse integral spectra for projections of individual proton energy bins were fit as described above, resulting in a light yield value for each bin for each trial. To obviate any potential issues resulting from the energy gradient of the neutron flux being large, the mean proton energy of all events in a given bin in a given trial was taken as the representative proton energy for that bin and trial, with the bin edges represented by asymmetrical error bars. The resulting set of light yield versus proton energy data points for each trial was stored on disk along with the associated errors. To reduce the Monte Carlo trials to a single set of data points and error bars, for a given energy bin, the mean value of the light yield and proton energy for all trials was taken as the light yield and proton energy, respectively. The standard deviation of the light yield values for all trials for a given energy bin was taken as the systematic uncertainty. Variance-covariance matrices were also generated from the Monte Carlo simulation and are supplied in Appendix~\ref{var-covar}. The variance of the proton energy values from trial to trial was neglected in the generation of the variance-covariance matrix. The result of the treatment of the systematic error for EJ-309 obtained using a 300~ns integration length is shown in Fig.~\ref{systematicResult}.


\section{Results and Discussion}
\label{results}

\begin{figure}
\centering
\includegraphics[width=0.8\textwidth]{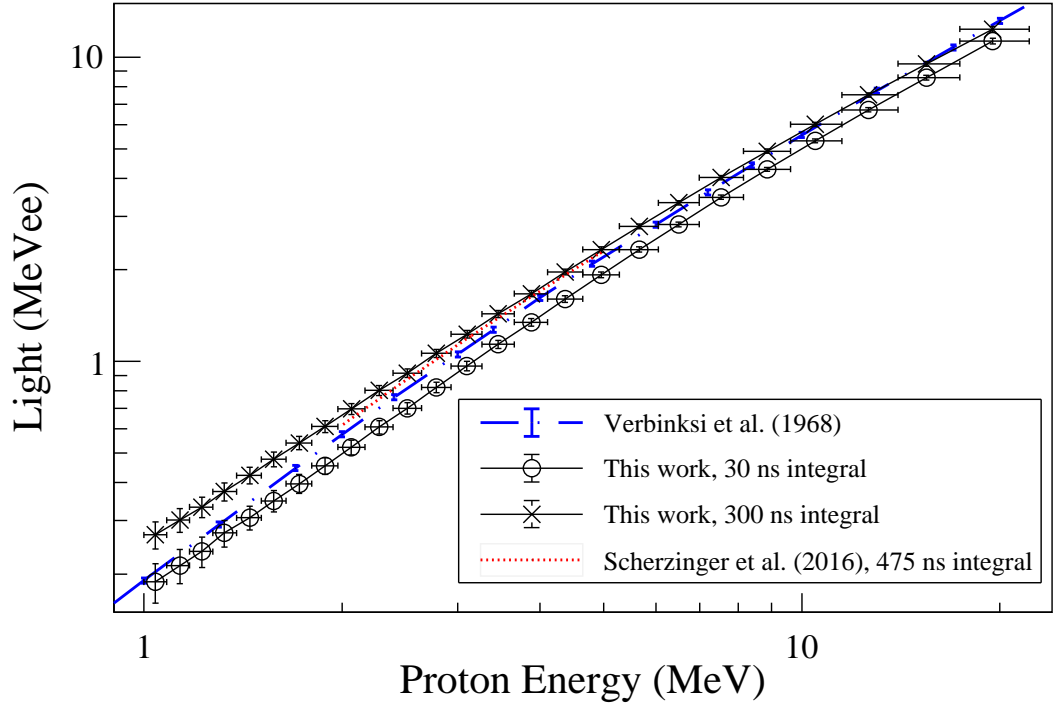}
\caption{EJ-301 proton light yield.}
\label{EJ301Result}
\end{figure}

\begin{figure}
	\centering
	\includegraphics[width=0.8\textwidth]{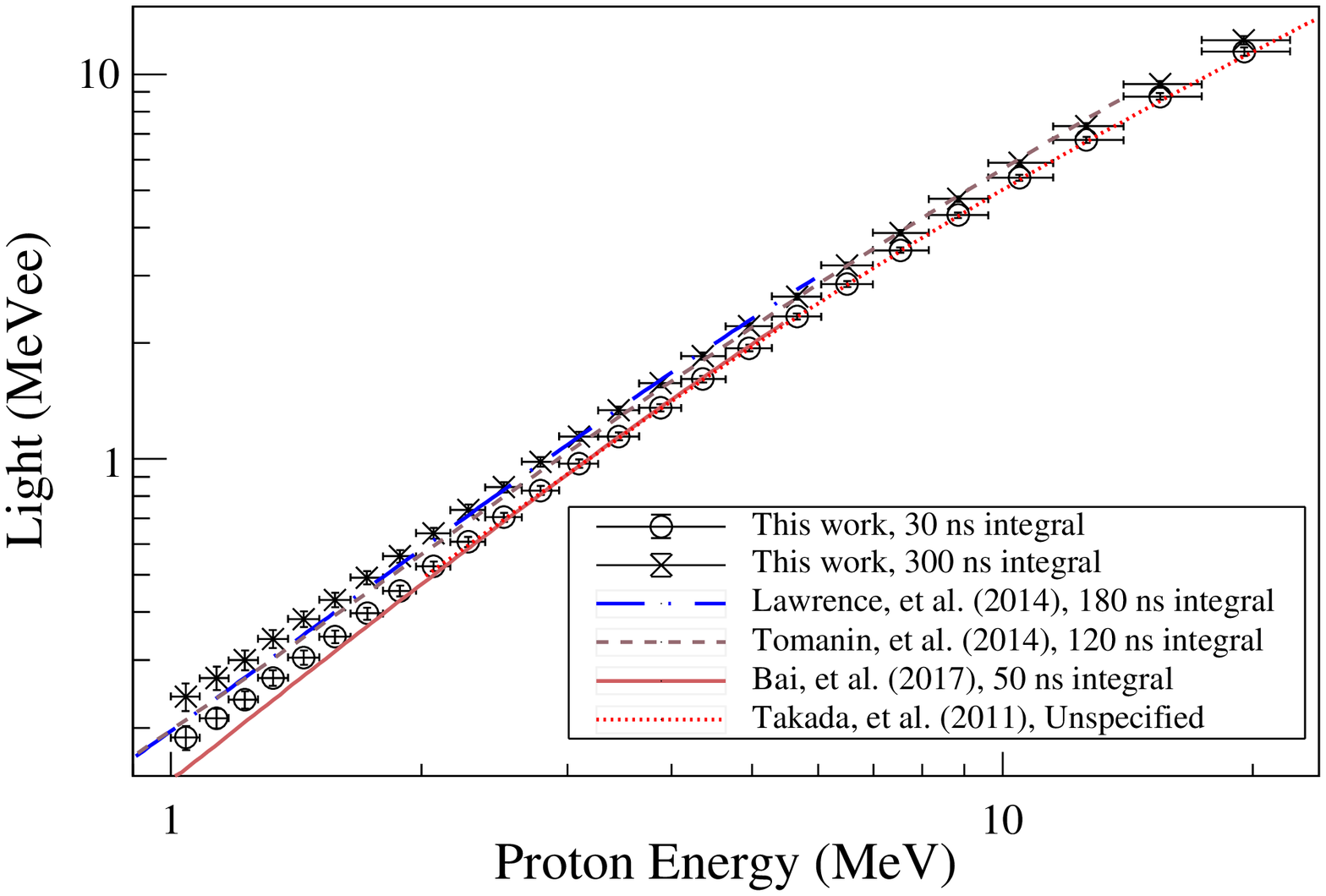}
	\caption{EJ-309 light yield from $1-20$~MeV proton recoil energy.}
	\label{EJ309ResultFull}
\end{figure}

The measured proton light yield of EJ-301 and EJ-309 are shown in Figs.~\ref{EJ301Result} and~\ref{EJ309ResultFull}, respectively. Both results are presented using 30~ns and 300~ns integration lengths and a summary table of the light yield data is provided in Appendix~\ref{LYDataPoints}. The EJ-301 results in Fig.~\ref{EJ301Result} are presented along with light yield measurements of the equivalent material, NE-213. The measurement from Verbinski et al.\ was converted to MeVee using the preferred conversion factor from Dietze and Klein.\ \cite{verbinski, DIETZE} The best fit parameter result from Scherzinger et al.\ is also shown for a 475~ns pulse integration length.\cite{scherzinger} The short and long integral measurements obtained using the double TOF technique differ significantly over the full energy range of the measurement. The long integral measurement is in agreement with literature values within estimated uncertainties above proton recoil energies of $\sim$3~MeV and systematically lies above the result of Verbinski et al.\ at lower energies. 

\begin{table*} 
\renewcommand{\arraystretch}{1.2}
\begin{tabular}{cccc}
\hline
Author & Method & Shape & Size \\
\hline
This work & double TOF & right cylinder &  5.08 cm len.\ $\times$ 5.08 cm dia.\ \\
Takada et al.\ \cite{takada} & direct & right cylinder & 12.17 cm len.\ $\times$ 12.17 cm dia.\ \\
Bai et al.\ \cite{bai} & edge characterization with trial response & right cylinder & 5.1 cm len. $\times$ 5.1 cm dia.\ \\
Lawrence et al.\ \cite{lawrence} & edge characterization & right cylinder & 7.62 cm len. $\times$ 7.62 cm dia.\ \\
Tomanin et al.\ \cite{tomanin} & trial response & cubic & 10 cm cu.\ \\
\hline
\end{tabular}
\caption{Summary of EJ-309 light yield measurements and methods, including scintillator cell shape and size.}
\label{cells}
\end{table*}

The short and long integral measurements of the EJ-309 light yield using the double TOF technique are presented in Fig.~\ref{EJ309ResultFull}. These are compared against measurements from the literature obtained using a variety of approaches, summarized in Table~\ref{cells}. Takada et al.\ used a direct method, where protons were impinged directly into an EJ-309 scintillator cell, with an unspecified integration length.~\cite{takada} Bai et al.\ used a short integration length with an edge characterization technique that incorporated a Monte Carlo simulation to obtain the final light yield relation through comparison between measured and simulated pulse integral spectra.\cite{bai} Lawrence et al.\ and Tomanin et al.\ used edge characterization and a Monte Carlo trial response, respectively, with long integration lengths.\cite{lawrence, tomanin} 

\begin{figure}
	\centering
	\includegraphics[width=0.8\textwidth]{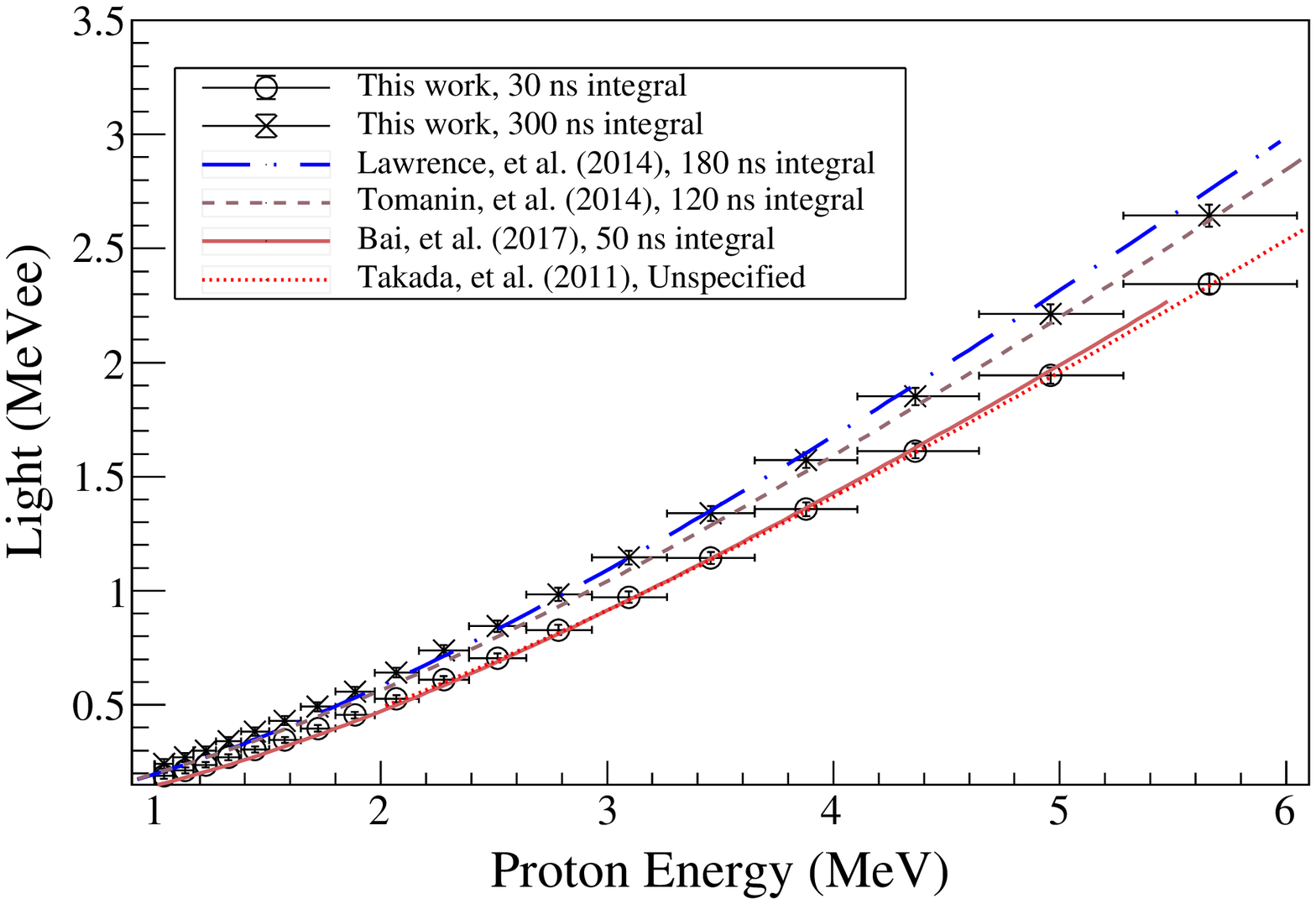}
	\caption{EJ-309 light yield from $1-6$~MeV proton recoil energy.}
	\label{EJ309-LowELin}
\end{figure}

Integration length plays a key role in the extracted proton light yield. As shown in Fig.~\ref{EJ309-LowELin}, the EJ-309 light yield measurements cluster in two groups reflective of the integration lengths used in the pulse processing chains, where the double TOF results for both the short and long pulse integration windows are in general agreement with literature data using similar pulse integration lengths. As the scintillation light resulting from proton energy deposition in the medium has a relatively larger delayed emission component compared to $\gamma$ rays, the proton light yield data with short integration lengths lies systematically below long integration length data. The observed systematic difference between short and long integration length light yield results is not merely due to the inclusion of more light in the pulse integral of the proton data, as longer integration lengths also increase the amount of light in the $\gamma$-ray calibration data. As such, changes in integration length change the electron-equivalent energy scale to be representative of different numbers of photons for a given electron energy. Further, as the proton pulse shape (i.e., delayed-to-prompt ratio) varies as a function of proton energy, discrepancies between short and long integration lengths are not described using a simple multiplicative factor. For example, for the double TOF EJ-309 light yield data, the percent difference between the light yield obtained using the 30~ns and 300~ns integration lengths varies from approximately $6-22$\% over the full energy range of the measurement. 

This comparison against literature measurements does not support the notion that proton light yield (in electron equivalent light units) is a detector-specific quantity.\cite{scherzinger, tomanin, enqvist} The EJ-309 proton light yield measurements obtained in this work for proton recoil energies above approximately 1.5~MeV bound the literature measurements independent of the shape or size of the scintillation material, the readout system, or the method used to characterize the material. Remaining differences among the data in each pulse integration length group may be originating from the lack of standardization in MeVee calibration procedures and material variations (e.g., fluor concentration, dissolved oxygen content, etc.)\ amongst scintillator samples from different lots. 


\section{Summary}
\label{conc}

A new model-independent method for measuring the proton light yield relation of organic scintillators has been developed that extends the indirect measurement technique for use with a broad spectrum neutron source. Using a double TOF technique and a high-flux, pulsed neutron beam, the proton light yield of EJ-301 and EJ-309 organic liquid scintillators was measured over a continuous proton recoil energy range from 1 to 20~MeV in a single measurement. The results were calibrated in an electron equivalent light unit using $\gamma$-ray sources via a $\chi^2$ minimization between simulations of the electron energy deposition and the measured distributions. The systematic uncertainty in the light yield measurement was characterized, including an evaluation of the covariance of the light yield data. The existing discrepancies in the literature for the proton light yield of EJ-309 are largely explained by differences in the integration lengths used in the pulse processing chains. This work provides a new avenue of inquiry for light yield measurements for basic and applied science and sheds light on the rationale for the variations in the measured EJ-309 light yield reported in the literature.

\appendix
\section{Light Yield Variance-Covariance Matrix}
\label{var-covar}

\begin{figure*}[t!]
	\centering
	\begin{subfigure}{0.5\textwidth}
		\centering
	    \includegraphics[width=0.97\textwidth]{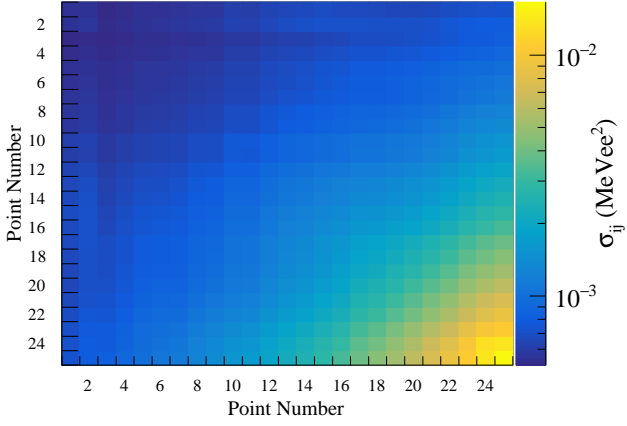}
        \caption{EJ-301 light yield covariance matrix.}
        \label{EJ301Covariance}
	\end{subfigure}%
	~ 
	\begin{subfigure}{0.5\textwidth}
		\centering
        \includegraphics[width=0.97\textwidth]{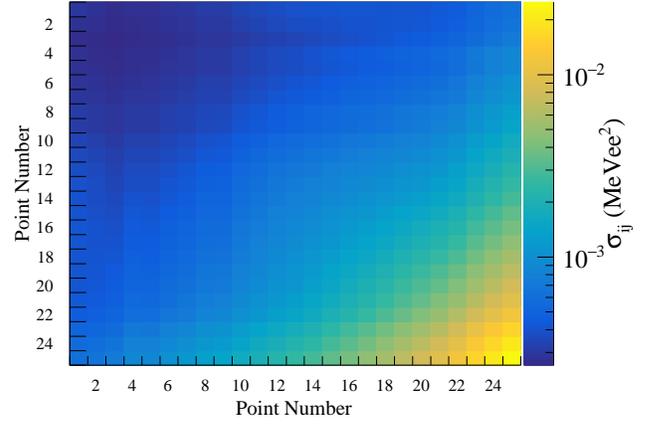}
        \caption{EJ-309 light yield covariance matrix.}
        \label{EJ309Covariance}
	\end{subfigure}

	\begin{subfigure}{0.5\textwidth}
	  \centering
      \includegraphics[width=0.97\textwidth]{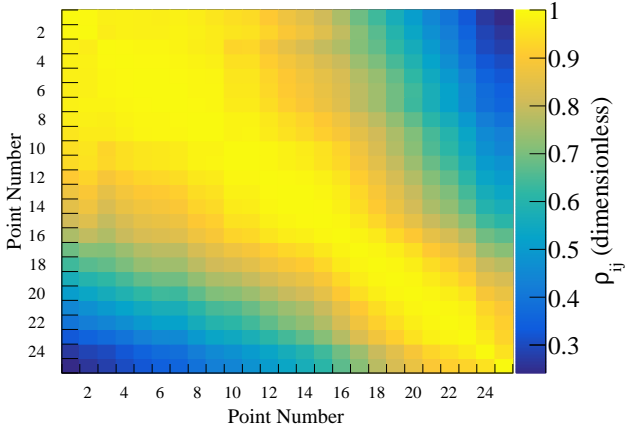}
      \caption{EJ-301 light yield correlation matrix.}
      \label{EJ301Correlation}
    \end{subfigure}%
~ 
\begin{subfigure}{0.5\textwidth}
	\centering
    \includegraphics[width=0.97\textwidth]{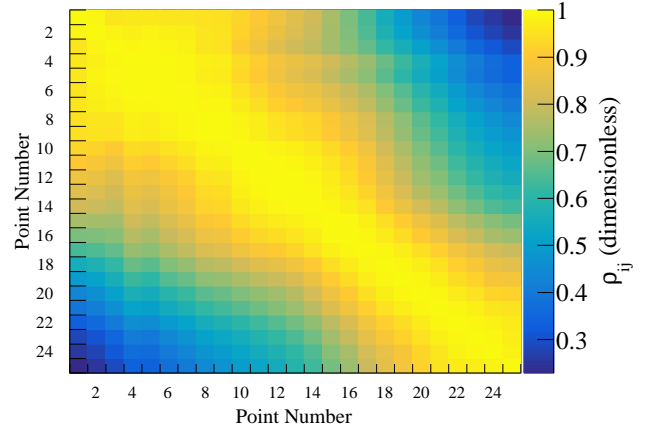}
    \caption{EJ-309 light yield correlation matrix.}
    \label{EJ309Correlation}
\end{subfigure}
	\caption{Covariance and correlation matrices for EJ-301 and EJ-309 Monte Carlo estimation of the systematic uncertainties. The $x$- and $y$-axes correspond to the data points given in Table~\ref{resultsTable}, with increasing proton recoil energy, for the 300~ns integration lengths. Reprinted from J.A. Brown, \textit{A Double Time of Flight Method for Measuring Proton Light Yield,} PhD thesis, University of California, Berkeley, December 2017.}
\end{figure*}

The covariance $\sigma_{ij}$ between two data points, $i$ and $j$, is given by: 
\begin{equation}
\label{covariance}
\sigma_{ij} = \frac{1}{N} \sum_{n=0}^{N}(L_i^{(n)}-\mu_i)(L_j^{(n)}-\mu_j),
\end{equation} 
where $N$ is the number of Monte Carlo iterations, $L_i^{(n)}$ is the calculated light yield for the $i^{th}$ data point in the $n^{th}$ iteration, and $\mu_i$ is the mean light yield for the $i^{th}$ data point. The variance-covariance matrices for EJ-301 and EJ-309 light yield data obtained using 300~ns integration lengths are shown in Fig.~\ref{EJ301Covariance} and \ref{EJ309Covariance}, respectively. The correlation $\rho_{ij}$ between two data points, $i$ and $j$, is given by: 
\begin{equation}
\label{correlation}
\rho_{ij} = \frac{\sigma_{ij}}{\sigma_i\sigma_j},
\end{equation} 
where $\sigma_i$ is the standard deviation for the $i^{th}$ data point. The correlation matrices for EJ-301 and EJ-309 light yield data obtained using 300~ns integration lengths are shown in Fig.~\ref{EJ301Correlation} and \ref{EJ309Correlation}, respectively.

\section{Light Yield Data}
\label{LYDataPoints}
\label{appendix}

The measured light yield data for EJ-301 and EJ-309 using 30~ns and 300~ns integration lengths are summarized in Table~\ref{resultsTable}. The asymmetric proton recoil energy error bars are reflective of the non-uniform distribution of proton energies within a given bin.

\begin{table*}
	\centering
	\renewcommand{\arraystretch}{1.2}
	\setlength{\tabcolsep}{10pt}
	\begin{tabular}{ccc|ccc}
		\hline
		\multicolumn{3}{c|}{EJ-301} & \multicolumn{3}{c}{EJ-309}   \\
		Proton recoil   & \multicolumn{2}{c|}{Light Yield (MeVee)} & Proton recoil  & \multicolumn{2}{c}{Light Yield (MeVee)}  \\
		energy (MeV)&  30 ns & 300 ns & energy (MeV)& 30 ns & 300 ns \\
\hline
1.04  $^{+0.04}_{-0.04}$  & 0.188 $\pm$ 0.013 & 0.241  $\pm$ 0.020 & 1.04  $^{+0.04}_{-0.04}$ & 0.187 $\pm$ 0.028 & 0.269 $\pm$ 0.027 \\
1.14  $^{+0.04}_{-0.05}$  & 0.212 $\pm$ 0.013 & 0.269  $\pm$ 0.019 & 1.13  $^{+0.04}_{-0.05}$ & 0.212 $\pm$ 0.027 & 0.301 $\pm$ 0.027 \\
1.23  $^{+0.05}_{-0.05}$  & 0.236 $\pm$ 0.012 & 0.300  $\pm$ 0.018 & 1.23  $^{+0.05}_{-0.05}$ & 0.236 $\pm$ 0.027 & 0.332 $\pm$ 0.025 \\
1.33  $^{+0.06}_{-0.05}$  & 0.269 $\pm$ 0.012 & 0.341  $\pm$ 0.018 & 1.32  $^{+0.06}_{-0.05}$ & 0.271 $\pm$ 0.027 & 0.374 $\pm$ 0.026 \\
1.44  $^{+0.06}_{-0.06}$  & 0.304 $\pm$ 0.012 & 0.384  $\pm$ 0.018 & 1.45  $^{+0.06}_{-0.06}$ & 0.305 $\pm$ 0.027 & 0.422 $\pm$ 0.026 \\
1.58  $^{+0.07}_{-0.07}$  & 0.345 $\pm$ 0.013 & 0.431  $\pm$ 0.019 & 1.58  $^{+0.07}_{-0.07}$ & 0.347 $\pm$ 0.027 & 0.477 $\pm$ 0.026 \\
1.72  $^{+0.08}_{-0.08}$  & 0.396 $\pm$ 0.014 & 0.491  $\pm$ 0.020 & 1.72  $^{+0.08}_{-0.08}$ & 0.395 $\pm$ 0.028 & 0.539 $\pm$ 0.027 \\
1.89  $^{+0.09}_{-0.09}$  & 0.454 $\pm$ 0.015 & 0.559  $\pm$ 0.021 & 1.89  $^{+0.09}_{-0.09}$ & 0.453 $\pm$ 0.028 & 0.611 $\pm$ 0.028 \\
2.07  $^{+0.10}_{-0.10}$  & 0.525 $\pm$ 0.015 & 0.641  $\pm$ 0.021 & 2.07  $^{+0.10}_{-0.09}$ & 0.521 $\pm$ 0.029 & 0.699 $\pm$ 0.028 \\
2.28  $^{+0.11}_{-0.11}$  & 0.608 $\pm$ 0.018 & 0.737  $\pm$ 0.024 & 2.28  $^{+0.11}_{-0.11}$ & 0.608 $\pm$ 0.030 & 0.802 $\pm$ 0.031 \\
2.52  $^{+0.13}_{-0.13}$  & 0.705 $\pm$ 0.020 & 0.845  $\pm$ 0.026 & 2.51  $^{+0.13}_{-0.12}$ & 0.701 $\pm$ 0.030 & 0.916 $\pm$ 0.031 \\
2.78  $^{+0.15}_{-0.14}$  & 0.827 $\pm$ 0.022 & 0.984  $\pm$ 0.028 & 2.78  $^{+0.15}_{-0.14}$ & 0.822 $\pm$ 0.032 & 1.062 $\pm$ 0.034 \\
3.10  $^{+0.17}_{-0.16}$  & 0.971 $\pm$ 0.024 & 1.145  $\pm$ 0.030 & 3.10  $^{+0.17}_{-0.17}$ & 0.964 $\pm$ 0.033 & 1.231 $\pm$ 0.036 \\
3.46  $^{+0.20}_{-0.19}$  & 1.144 $\pm$ 0.026 & 1.339  $\pm$ 0.032 & 3.45  $^{+0.20}_{-0.19}$ & 1.136 $\pm$ 0.035 & 1.432 $\pm$ 0.038 \\
3.88  $^{+0.23}_{-0.23}$  & 1.357 $\pm$ 0.029 & 1.573  $\pm$ 0.035 & 3.88  $^{+0.23}_{-0.23}$ & 1.344 $\pm$ 0.037 & 1.670 $\pm$ 0.040 \\
4.36  $^{+0.28}_{-0.25}$  & 1.611 $\pm$ 0.033 & 1.852  $\pm$ 0.038 & 4.36  $^{+0.28}_{-0.26}$ & 1.603 $\pm$ 0.038 & 1.966 $\pm$ 0.043 \\
4.96  $^{+0.32}_{-0.32}$  & 1.942 $\pm$ 0.036 & 2.213  $\pm$ 0.042 & 4.96  $^{+0.33}_{-0.31}$ & 1.926 $\pm$ 0.042 & 2.331 $\pm$ 0.048 \\
5.66  $^{+0.39}_{-0.38}$  & 2.344 $\pm$ 0.042 & 2.644  $\pm$ 0.048 & 5.66  $^{+0.39}_{-0.38}$ & 2.331 $\pm$ 0.045 & 2.780 $\pm$ 0.051 \\
6.50  $^{+0.49}_{-0.45}$  & 2.848 $\pm$ 0.049 & 3.188  $\pm$ 0.055 & 6.50  $^{+0.49}_{-0.45}$ & 2.822 $\pm$ 0.049 & 3.323 $\pm$ 0.057 \\
7.54  $^{+0.61}_{-0.55}$  & 3.488 $\pm$ 0.057 & 3.872  $\pm$ 0.064 & 7.54  $^{+0.61}_{-0.55}$ & 3.460 $\pm$ 0.056 & 4.023 $\pm$ 0.066 \\
8.84  $^{+0.77}_{-0.69}$  & 4.304 $\pm$ 0.069 & 4.741  $\pm$ 0.076 & 8.86  $^{+0.76}_{-0.70}$ & 4.279 $\pm$ 0.066 & 4.910 $\pm$ 0.078 \\
10.48 $^{+1.02}_{-0.87}$  & 5.387 $\pm$ 0.089 & 5.890  $\pm$ 0.097 & 10.48 $^{+1.03}_{-0.86}$ & 5.306 $\pm$ 0.081 & 6.013 $\pm$ 0.095 \\
12.62 $^{+1.37}_{-1.12}$  & 6.75  $\pm$0.13   & 7.33   $\pm$ 0.13  & 12.63 $^{+1.36}_{-1.13}$ & 6.71  $\pm$ 0.11  & 7.51  $\pm$ 0.13  \\
15.47 $^{+1.90}_{-1.48}$  & 8.76  $\pm$0.18   & 9.43   $\pm$ 0.19  & 15.46 $^{+1.91}_{-1.47}$ & 8.57  $\pm$ 0.16  & 9.51  $\pm$ 0.19  \\
19.51 $^{+2.64}_{-2.15}$  & 11.47 $\pm$ 0.28  & 12.27  $\pm$ 0.30  & 19.50 $^{+2.67}_{-2.12}$ & 11.29 $\pm$ 0.24  & 12.37 $\pm$ 0.27  \\
\hline
\end{tabular}
	\caption{Light yield data for EJ-309 and EJ-301 obtained using 30~ns and 300~ns integration lengths.}
	\label{resultsTable}
\end{table*}

\begin{acknowledgments}
The authors would like to thank the 88-Inch Cyclotron operations and facilities staff for their help in performing this experiment and recognize the contributions of Andrew Voyles, Ethan Boado, Will Kable, and J{\o}rgen Midtb{\o} in the data collection. This work was performed under the auspices of the U.S. Department of Energy National Nuclear Security Administration by Lawrence Livermore National Laboratory under Contract DE-AC52-07NA27344, Lawrence Berkeley National Laboratory under Contract DE-AC02-05CH11231, and through the Nuclear Science and Security Consortium under Award Nos. DE-NA0000979 and DE-NA0003100. Sandia National Laboratories is a multimission laboratory managed and operated by National Technology and Engineering Solutions of Sandia LLC, a wholly owned subsidiary of Honeywell International Inc., for the U.S. Department of Energy's National Nuclear Security Administration under contract DE-NA0003525.
\end{acknowledgments}
\clearpage

\bibliography{./SandiaPreprint}
\bibliographystyle{unsrtnat}

\end{document}